\documentclass{llncs}
\usepackage[T1]{fontenc}
\usepackage{graphicx}
\usepackage{todonotes}
\usepackage{hyperref}
\usepackage[capitalise]{cleveref}
\usepackage{cite}
\usepackage{booktabs}
\usepackage{dcolumn}
\usepackage{tikz}
\usepackage{pgfplots}
\usepackage[font=small,labelfont=bf]{caption}
\usepackage[font=small,labelfont=small]{subcaption}
\usepackage{pifont}

\title{A Simple Pipeline for Orthogonal Graph Drawing}

\author{}
\institute{}

\author{Tim Hegemann\orcidID{0009-0008-4770-3391} \and
  Alexander Wolff\orcidID{0000-0001-5872-718X}}
\authorrunning{T.~Hegemann and A.~Wolff}
\institute{Universit\"at W\"urzburg, W\"urzburg, Germany
  \\\email{hegemann@informatik.uni-wuerzburg.de}}

\usetikzlibrary{plotmarks}
\pgfplotsset{compat=1.17}

\graphicspath{{figures/}}
\renewcommand{\orcidID}[1]{\href{https://orcid.org/#1}{\includegraphics[scale=.03]{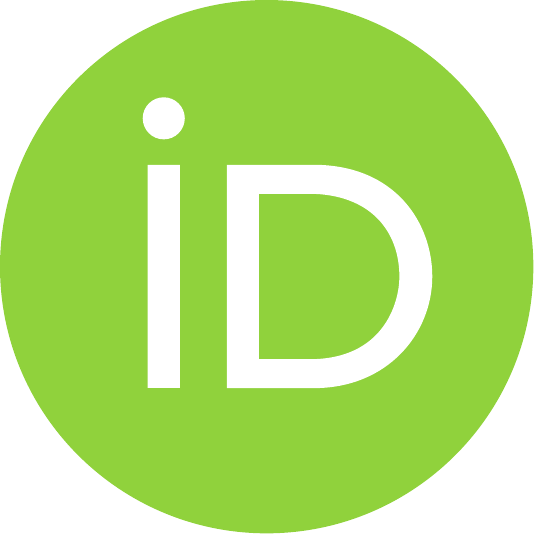}}}

\definecolor{PKdarkblue}{rgb}{0.121,0.47,0.705}
\definecolor{PKdarkred}{rgb}{0.89 0.102 0.109}
\definecolor{PKdarkgreen}{rgb}{0.2 0.627 0.172}
\definecolor{PKdarkorange}{rgb}{1 0.498 0}
\definecolor{PKdarkpurple}{rgb}{0.415 0.239 0.603}
\definecolor{PKlightgray}{rgb}{0.8 0.8 0.8}

\let\emph\relax
\DeclareTextFontCommand{\emph}{\color{PKdarkblue}\em}

\newcommand*\deltamin{\ensuremath{\delta_{\min}}}

\begin{document}

\maketitle

\begin{abstract}
  Orthogonal graph drawing has many applications, e.g., for laying out
  UML diagrams or cableplans.  In this paper, we present a new
  pipeline that draws multigraphs orthogonally, using few bends, few
  crossings, and small area.  Our pipeline computes an initial graph
  layout, then removes overlaps between the rectangular nodes, routes
  the edges, orders the edges, and nudges them, that is, moves edge
  segments in order to balance the inter-edge distances.  Our pipeline
  is flexible and integrates well with existing approaches.  Our main
  contribution is (i)~an effective edge-nudging algorithm that is
  based on linear programming, (ii)~a selection of simple
  algorithms that together produce competitive results, and
  (iii)~an extensive experimental comparison of our pipeline with
  existing approaches using standard benchmark sets and metrics.

\end{abstract}

\section{Introduction}
\label{sec:intro}

Due to its many applications, the orthogonal drawing style has been
studied extensively in Graph Drawing.  One of the milestones in the
development of efficient algorithms for this domain was Tammasia's
Topology--Shape--Metric framework \cite{t-eggmn-SJC87} which showed
that embedded planar graphs with a vertex degree of at most~4 can be
laid out efficiently, using the minimum number of bends.  The
restriction to degree~4 comes from the fact that the framework
represents vertices by (grid) points.  For practical purposes,
however, the restriction to (embedded) planar graphs of constant
degree is prohibitive.  This triggered many practical approaches to
orthogonal graph drawing.
For example, the three-phase method of Biedl et al.~\cite{Biedl2000}
draws \emph{normalized} graphs (that is, graphs without self-loops and
leaves) with vertices in general position (that is, on different grid
lines) with ``small'' vertex boxes on an quadratic size grid using at
most one bend per edge.  For compaction, Biedl et al.\ referred to
Lengauer's book \cite{lengauer1990} on VLSI layout, which is also
relevant for orthogonal graph drawing.
Building on earlier work
\cite{dwyer2006,dkm-ipsep-TVCG06,wybrow2010,rueegg2014}, Kieffer et
al.~\cite{kieffer2016} introduced \textsc{Hola} (``Human-like
Orthogonal Network Layout''), a multi-step approach for drawing graphs
orthogonally.  They partition the input graph and use different layout
strategies for different parts: stress minimization for the graph core
and specialized code for tree-like subgraphs.  %

Schulze et al.~\cite{ssh-dlgpc-JVLC14} presented the orthogonal
graph drawing library \textsc{Kieler}~\cite{kieler20} that took special care of
so-called \emph{port constraints}.  They allow the user to specify on
which side of a vertex box an edge must be attached, which is
important, for example, for UML diagrams.  Zink et
al.~\cite{zwbw-ldugg-CGTA22} presented the \textsc{Praline} library,
which generalizes port constraints by
introducing port groups and port pairings, which are useful for
drawing cableplans.  Both the approaches of Schulze et al.\ and Zink
et al.\ arrange nodes on layers and use the framework of Sugiyama et
al.~\cite{Sugiyama1981} for layered graph drawing (although they do
not assume input graphs to be directed), among others, in order to
reduce edge crossings.

Whereas most graph drawing algorithms place labels into vertex boxes,
Binucci et al.~\cite{binucci2005} %
also incorporated {\em edge labels}.
Using mixed integer programming, they can draw
sparse graphs with vertex and edge labels of up to 100 vertices.

We considered orthogonal graph layout in a third-party-funded project
with two industrial partners with different backgrounds.
One of them produces network management software; the other produces
software for drawing cableplans of complicated, highly configurable
machines.  Both asked for layouts that work well on mobile devices
such as tablet computers used by, for example, technicians who service harvesting
machines in the field.

Rather than a monolithic software package, they were interested in a highly
configurable, flexible pipeline whose parts can easily be exchanged in
order to meet the various needs of their customers.  Still, they
insisted on a number of basic requirements.  The drawings computed by
our algorithm must be orthogonal, that is, edges are drawn as
sequences of axis-aligned segments and vertices are represented by
non-overlapping boxes (i.e., axis-aligned rectangles).  Also, the
user must be able to specify a minimum object distance~\deltamin\ to
be respected by vertex boxes and edge segments.

In terms of quality, we agreed upon standard graph drawing criteria
such as few crossings, few bends, small area, good aspect ratio, small
total edge length, and small edge length variance.  With mobile
applications in mind, using small area becomes our key objective.  However,
rather than an algorithm that excels in one of these metrics (and fails
badly in another), our partners were interested in allrounders that are
sufficiently fast and generally perform well.

\subsubsection{Our Contribution.}

We set up a layout pipeline with three variants.
For the first variant (\textsc{Force}), we use a force-directed layout
algorithm~\cite{fr-gdfdp-SPE91} to place the vertices of a given graph $G$
as points (ignoring their boxes).  Then we center the vertex boxes on these points.
If some of them overlap, we call an overlap removal algorithm of Nachmanson
et al.~\cite{nachmanson2017}.  Instead of these three steps, for the second
variant (\textsc{Hybrid1}), we used the vertex placement computed by \textsc{Praline}.
In both cases, we apply the following steps that we describe in detail
in \cref{sec:pipeline}. \textit{Port assignment:} We assign the endpoints
of the edges to the sides of the vertex boxes.  \textit{Routing graph construction:}
We construct an auxiliary grid-like graph~$H$.  \textit{Edge routing} and
\textit{path ordering:} We route (and order) the edges of~$G$ along the edges of~$H$.
Our path ordering is based on existing techniques~\cite{pupyrev2016,groeneveld89,noellenburg2010}.
Now each edge of~$G$ consists of a path of %
axis-aligned segments.
\textit{Edge nudging:} In this final step, we distribute the path segments so that
they partition the available space between the vertex boxes as evenly as possible.
As a third variant (\textsc{Hybrid2}), we initialize our pipeline with
both the vertex positions and the edge routing computed by \textsc{Praline} and
apply only the nudging step as a post-processing.

Our main contribution is (i)~the edge nudging step, (ii)~our simple and
flexible pipeline as a whole, and (iii)~an experimental comparison
with the state-of-the-art orthogonal layout libraries \textsc{Hola}\footnote{
  see \url{https://www.adaptagrams.org/}}
\cite{kieffer2016} and \textsc{Praline}\footnote{see
\url{https://github.com/j-zink-wuerzburg/praline}}\cite{zwbw-ldugg-CGTA22}
on two standard benchmark sets (see \cref{sec:experiments}).
\textsc{Praline} has already been compared with \textsc{Kieler}, and
performed similarly well or even slightly
better~\cite{zwbw-ldugg-CGTA22}.  \textsc{Hola} has been
compared to the \textit{orthogonal style} automatic layout of
\textsc{yFiles}\footnote{see \url{https://www.yworks.com/products/yfiles}}%
; \textsc{Hola} outperformed \textsc{yFiles}
almost universally in a user study with $89$~participants~\cite{kieffer2016}.

As it turns out, our pipeline proves to be a good allrounder that
performs well in many of the metrics mentioned above.  Due to careful
nudging, our pipeline yields very compact layouts, whereas its
competitors usually produce fewer crossings and bends.  Depending on
the variant, our pipeline takes slightly less or about twice as much
time than \textsc{Praline}.  It is much faster than \textsc{Hola}.

Our source code is available at %
\url{https://github.com/hegetim/wueortho}.

\section{Our Pipeline}
\label{sec:pipeline}

In order to fine-tune the layout for different requirements
(as discussed in \cref{sec:intro}), we designed our pipeline as a sequence
of mostly independent, interchangeable, and self-contained steps that we
describe in detail in this section.

The input for our pipeline is a multigraph $G$ with vertex set $V(G)$
of size $n$ and edge multiset~$E(G)$ of size $m$.  We explicitly
allow our graphs to have self-loops and handle them in the \textit{port
assignment} step.  Each vertex comes with a (textual) vertex label or
directly with a \emph{vertex box}, that is, an axis-aligned rectangle.
Given a text label, we compute a box that fits the label (in some standard font).

\begin{figure}[tb]
  \centering
  \includegraphics[page=1]{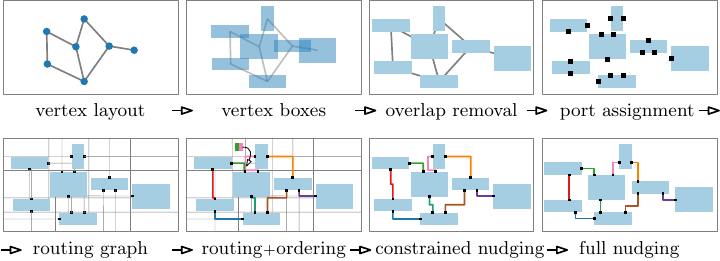}
  \caption{Our flexible pipeline of simple algorithms for orthogonal graph drawing.}
  \label{fig:pipeline}
\end{figure}

Our pipeline consists of several simple algorithms for specific
subproblems of orthogonal graph drawing; see \cref{fig:pipeline} for
an overview.  The main steps in our pipeline are vertex layout,
overlap removal, port assignment, construction of the routing graph,
edge routing, path ordering, and edge nudging.  Below we detail most of these
steps.  For the remaining, we use standard algorithms, see ``Pipeline
Variants'', \cref{sec:experiments}.

\subsubsection{Port Assignment.}

Each edge connects one or two vertices that are represented by rectangular boxes.
We call the start point and the end point of an edge its \emph{ports}.
Ports lie on the boundary of the vertex boxes that an edge connects.
Port positions can either be specified in the input, or they are set
by our pipeline as follows (before the edge routes are determined).

For each edge $uv$, we first determine the sides of the boxes of~$u$
and~$v$ on which we place the ports of~$uv$.  Let~$s_{uv}$ be the
straight-line segment that connects the centers of the boxes of~$u$
and~$v$.  We usually assign the ports to those box sides that
intersect~$s_{uv}$.  To avoid situations where an edge~$uv$
would have to be drawn as a Z-shape according to the rule above, we
adjust the rule as follows.  We split each side evenly into four pieces.
If~$s_{uv}$ intersects a side in the first or last piece of the side,
we instead reassign one of the two ports such that $uv$ can be drawn as an
L-shape that bends away from the barycenter of the vertex boxes' center points.
(We do not actually route $uv$ now, we just use geometry to place its ports.)
The order along a side of vertex $u$ is then determined by the circular
order of the segments of type $s_{uv}$, where $v$ is a neighbor of $u$.
Multi-edges require special care regarding their order within a side.
They are routed next to each other without crossings.  Self-loops
get neighboring ports assigned to the least populated side.
When all ports have been assigned to a box, we evenly distribute the
ports on each side.

\subsubsection{Routing Graph Construction.}

We route each edge of the input graph along an obstacle-avoiding path
in a \emph{routing graph} $H$ whose vertices are the ports and the
potential bend points of edge routes.  This graph forms a partial grid
with gaps around the vertex boxes.  A precise definition follows below.

\begin{figure}[tb]
  \centering
    \centering
    \includegraphics[page=2]{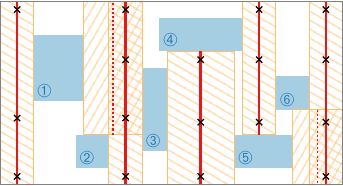}
    \caption{Routing graph construction: vertical channels (orange hatched)
    and their representatives (red).  Dotted representatives have been sorted
    out.  Note that ports are omitted and all representatives are the vertical
    center line of their channel.  Black crosses mark where horizontal
    representatives (not shown) intersect.  Those will host vertices in the
    final routing graph.}
    \label{fig:routing-graph}
\end{figure}

The intuition for our routing graph is that edges get routed through horizontal
and vertical channels.  We describe only \emph{vertical channels}.
\emph{Horizontal channels} are defined symmetrically.  For a pair of vertex
boxes $(u, v)$ where $v$ is entirely to the right of $u$, we define its
\emph{vertical channel} as the largest axis-aligned rectangle whose
vertical sides touch the right side of~$u$ and the left side of~$v$
and that is interior-disjoint from all vertex boxes~--
if such an empty rectangle exists for $(u, v)$.
For each box $u$ we keep only the channel to the right of $u$ that has
the smallest width.
For this step, we interpret the left and right boundaries
of the drawing as vertex boxes of zero width and infinite height.
In \cref{fig:routing-graph}, the orange boxes depict the vertical channels.
Note that some of these channels overlap.
We can find all channels in $\mathcal{O}(n \log n)$ time with a sweepline algorithm.

For each vertical channel, we define a vertical line segment that
spans the channel's entire height as its \emph{representative}.
If possible, we choose an appropriate line segment starting in a port.
Otherwise, we choose the center line of the channel.  As a further optimization,
we ignore every vertical channel $C$ that intersects another channel $C'$
such that the projection of $C$ on the y-axis is contained in that of $C'$.
See the dotted red segments in \cref{fig:routing-graph}.
We define representatives for horizontal channels symmetrically.
We add additional representatives for each remaining port.
We can find all representatives using the same sweepline algorithm
as above in $\mathcal{O}(m \log m)$ time, assuming $n \in \mathcal{O}(m)$.

The routing graph~$H$ has a vertex for each port and for each
intersection point between a vertical and a horizontal representative.
It has an edge between each pair of vertices that are consecutive
along a representative.  Let $M$ be the number of edges of~$H$.  Again
assuming $n \in \mathcal{O}(m)$, we have $M \in \mathcal{O}(m^2)$
since there are at most $4n$ channels (one per side for each vertex
box) and %
$2m$ ports.

\subsubsection{Edge Routing.}

In the next step, for each edge we connect its endpoints (which are
ports and thus vertices in the routing graph~$H$) by a shortest path
in~$H$.  Our approach for edge routing is similar to that of Wybrow et
al.~\cite{wybrow2010} except that they use the A$^{\!*}$ search for computing
shortest paths whereas we use Dijkstra's algorithm (for simplicity).
Recall that~$H$ is a partial grid graph.  Among all shortest paths,
we choose a bend-minimal one by augmenting the state used in Dijktra's
algorithm.  For each partial path ending in a node~$v$, in addition
to the length of the path, we store the number of bends and the direction
of entry when entering~$v$.
In the following, we refer to routed edges as paths.  Each such path can be
found in $\mathcal{O}(M \log M)$ time.

The edge routing algorithm of Wybrow et al.\ does not take crossings
into account. Therefore, as a post-processing, we apply an additional
\emph{crossing reduction step}.
Whenever two paths cross each other more than once,
we replace the section between the first and last shared vertex
in one path with the corresponding section of the other.  This ensures
that, eventually, every pair of edges crosses at most once.

\subsubsection{Path Ordering.}

We now know, for each edge of the routing graph~$H$, the set of edge
paths routed over this segment.  Where several edge paths
(forming an \emph{edge bundle})
share a segment, we determine a path order that minimizes crossings.

For non-orthogonal routing graphs, Pupyrev et al.~\cite{pupyrev2016}
observed that the problem of determining such a path order
is computationally equivalent to the metro-line crossing minimization (MLCM) problem:
Let $\widetilde{G}$ be a plane graph (such as the routing graph),
and let $P$ be a set of simple paths in~$\widetilde{G}$.
For each edge $e$ in $E(\widetilde{G})$, find an order of all paths that contain $e$
that minimizes the number of crossings among all pairs of paths.

\begin{figure}[tb]
  \begin{subfigure}{\textwidth} \centering
    \includegraphics[page=3]{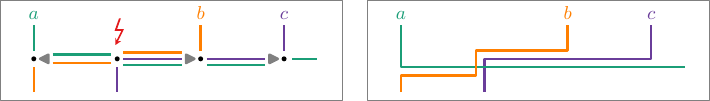}
    \subcaption{Edge paths ordered with the algorithm by Pupyrev et al.~\cite{pupyrev2016}.}
    \label{fig:path-order-a}
  \end{subfigure}\\[6pt]
  \begin{subfigure}{\textwidth} \centering
    \includegraphics[page=4]{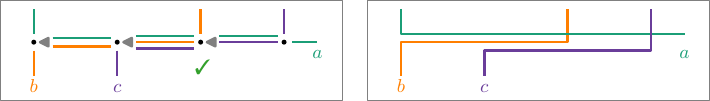}
    \subcaption{Edge paths ordered with our modified version of the algorithm.}
    \label{fig:path-order-b}
  \end{subfigure}
  \caption{Path ordering: The algorithm of Pupyrev et al.\ assigns each
    segment an arbitrary direction.  This can result
  in additional bends in the geometric realization (to the right).
  For orthogonal routing, we therefore preassign directions for
  horizontal and vertical segments.}
  \label{fig:path-order}
\end{figure}

We use the algorithm of Pupyrev et al.~\cite{pupyrev2016}
for our case where all edges are incident to unique ports (which makes
MLCM efficiently solvable).
When sorting paths in an edge bundle, for each pair of paths, they consider
the directions the paths take after leaving their common subpath at a
\emph{fork vertex}.  In \cref{fig:path-order-a},
for example, the leftmost vertex is a fork vertex for paths~$a$ and $b$.
With $a$ leaving to the top, it will be ordered above of $b$.
For each segment (i.e., edge in~$H$) such an ordering of paths has to be
found.  They fix an arbitrary direction that determines where to look for either
a fork vertex or a segment that has already been processed (see the gray
arrowheads in~\cref{fig:path-order}).  Crossings are unavoidable if the path
orders at the start and end of the common subpath differ.
Pupyrev et al.\ process the segments in arbitrary order.  Still, they introduce
at most one crossing for each pair of paths in an edge bundle and only if such
a crossing is unavoidable.
For example, paths $a$ and $b$ in \cref{fig:path-order-a} have an unavoidable
crossing.

In our orthogonal setting, if two paths change order between two adjacent
edges of~$H$ (see the red lightning in \cref{fig:path-order-a} (left)),
then we have to introduce two additional bends in a Z-shaped fashion in
their geometric realization (see \cref{fig:path-order-a} (right)).
In order to avoid such situations, we preassign directions for all segments
based on their orientation (left for horizontal segments and down for
vertical ones; see \cref{fig:path-order-b}).
Crossings now happen only where the orientation of segments changes
(i.e., at bend points of paths).  Such crossings can be realized
without additional edge bends; see the crossing at the green checkmark in
\cref{fig:path-order-b}.  Therefore, before the next step,
we join consecutive collinear segments of the same path.
Then, in any path, the orientation of the segments alternates.

Applied to the graph~$H$, our modification of the algorithm of
Pupyrev et al.\ runs in $\mathcal{O}(M k \log m)$ time, where $k$ is
the total number of segments in all paths.

\subsubsection{Edge Nudging.}

In the last step of our pipeline, we aim to balance the distances
between the path segments within their channels.
Our algorithm has two modes called \emph{constrained nudging}, when
vertex and port positions must not be altered, and \emph{full nudging}, when
a minimum distance between path segments and vertex boxes must be met,
but boxes can be moved and, if necessary, enlarged.
Both modes use basic linear programming (LP) to optimize segment distances.
They process horizontal and vertical distances independently.

\begin{figure}[tb]
  \centering
  \begin{subfigure}[b]{.48\linewidth}
    \centering
    \includegraphics[page=1]{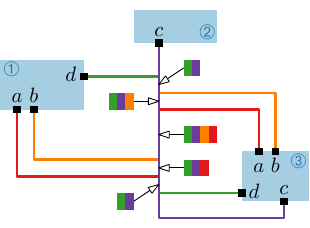}
    \subcaption{edge order indicated by colored boxes}
    \label{fig:nudging-a}
  \end{subfigure}
  \hfill
  \begin{subfigure}[b]{.48\linewidth}
    \centering
    \includegraphics[page=2]{nudging.pdf}
    \subcaption{defining the constraint graph}
    \label{fig:nudging-b}
  \end{subfigure}
  \caption{Edge nudging: Given an edge order on the vertical edge
    segments, we define horizontal separation constraints between vertical
    segments, left and right borders of vertex boxes, and two dummy segments (black bars).
    In (b) segments are partially nudged for readability.}
  \label{fig:nudging-ab}
\end{figure}

We now describe the horizontal pass.  The vertical pass works symmetrically.
First, we determine the horizontal order $\chi$ of all vertical path segments,
the left and right borders of all vertex boxes, and the two vertical
sides of a (slightly enlarged) bounding box of our instance
(see the black bars in \cref{fig:nudging-b}).
The order of the objects in $\chi$ is determined by their x-coordinate.
The two dummy segments are the first and last elements of~$\chi$.
It remains to define the order of objects with identical x-coordinate.
We assume non-intersecting, non-touching vertex boxes.
Where path segments overlap, the path order determined in the previous
section applies; see the colored boxes in \cref{fig:nudging-a}.
Right (left) borders of vertex boxes are inserted into $\chi$
before (after) any path segment with the same x-coordinate.
The order of non-overlapping path segments with the same x-coordinate
is arbitrary.

Given $\chi$, we define the \emph{constraint graph}~$G_\chi$,
the directed acyclic graph that has a vertex for each object as defined above,
and an arc from object~$u$ to object~$v$ if the vertical dimensions
of these objects overlap,~$u$ comes before~$v$ in~$\chi$, and there is no other
vertically overlapping object in between.
If this is the case, $u$ will be drawn to the left of~$v$.
Edges of this constraint graph will yield \emph{separation
constraints} of the form $u_x + \delta \le v_x$ in the LP,
where $u_x$ and $v_x$ are the x-coordinates of $u$ and $v$, respectively,
and $\delta$ is either a non-negative variable or the user-defined minimum
distance \deltamin{} between them.

\begin{figure}[tb]
  \centering
  \begin{subfigure}[t]{.49\linewidth}
    \centering
    \includegraphics[page=3]{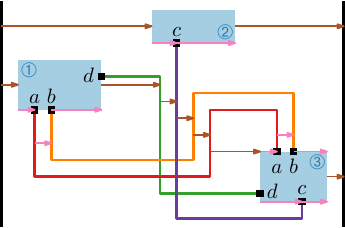}
    \subcaption{without transitive constraints}
    \label{fig:nudging-c}
  \end{subfigure}
  \hfill
  \begin{subfigure}[t]{.49\linewidth}
    \centering
    \includegraphics[page=4]{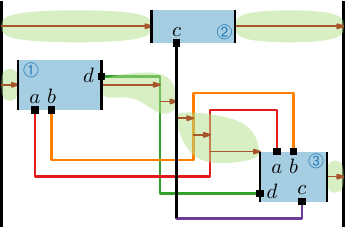}
    \subcaption{components of the constraint graph}
    \label{fig:nudging-d}
  \end{subfigure}
  \caption{Common steps of edge nudging:
  (a) After removing transitive arcs, pink arcs remain between unmovable objects.
      Brown arcs get distance variables.
  (b) The constraint graph is split at barriers (black bars).
      All constraints from arcs in the same component share their distance variables.}
  \label{fig:nudging-cd}
\end{figure}

Let $N=4n+m+b$ be the number of sides of the vertex boxes plus the number of edge
segments (i.e., the number of edges plus the number of bends).
Then the constraint graph can be constructed in $\mathcal{O}(N \log N)$ time,
using a sweepline algorithm of Dwyer et al.~\cite{dwyer2006}.
They showed that the number of edges in the constraint graph is linear in $N$
and that the separation constraints derived from the edges of $G_\chi$
guarantee a horizontally overlap-free drawing.
Next, we decide which constraints share the same \emph{distance
  variables} of type~$\delta$, which we will then maximize.
Wider channels with few segments allow for larger gaps than small crowded channels.
To obtain a balanced solution, we need to avoid situations
where two distance variables work against each other.

To identify preferably small sets of constraints that share the same distance
variable, we apply the following operations to the constraint graph.
We remove all transitive arcs, i.e., arcs $uw$ where also arcs $uv$
and $vw$ exist in the graph.  Constraints from these arcs are redundant.
We remove all arcs between objects that do not move in constrained nudging mode,
that is, the sides of vertex boxes and edge segments incident to ports;
see the purple arrows in \cref{fig:nudging-c}.  Then, the graph is split into
components (the green areas in \cref{fig:nudging-d}) that are confined by
unmovable objects or by dummy segments (the big black bars in
\cref{fig:nudging-d}).  All constraints derived from arcs of the same component
get the same distance variable.

\begin{figure}[tb]
  \centering
  \begin{subfigure}[b]{.49\linewidth}
    \centering
    \includegraphics[page=5]{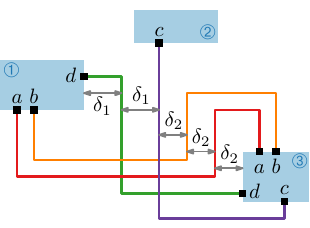}
    \subcaption{constrained nudging}
    \label{fig:nudging-e}
  \end{subfigure}
  \hfill
  \begin{subfigure}[b]{.49\linewidth}
    \centering
    \includegraphics[page=6]{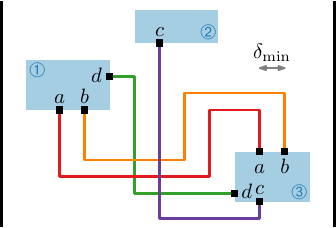}
    \subcaption{full nudging with distance \deltamin}
    \label{fig:nudging-f}
  \end{subfigure}
  \caption{Results of a horizontal nudging phase:
    (a) optimization nudges objects apart,
    (b) in full nudging mode,
    objects must maintain a given minimum distance \deltamin.
    Note that vertex \ding{194} has been slightly enlarged to make
    room for the ports of edges $a$ and $b$.}
  \label{fig:nudging-ef}
\end{figure}

In constrained nudging mode, for each movable or dummy segment,
we replace its position by a \emph{position variable} in all related constraints.
Finally, our LP minimizes %
\[ |W| (\omega - \alpha) - \sum_{\delta \in W} \delta, \]
where $\alpha$ and $\omega$ are the position variables of the left and right
dummy segments, respectively, and $W$ is the set of distance variables.
The factor $|W|$ is required to prevent the constraints involving distance
variables from pushing the dummy segments towards infinity.  The result is
shown in \cref{fig:nudging-e}.  Objects are separated with space between them
equivalent to at least the values of the respective distance variables.

In full nudging mode we allow both, the port segments and the borders
of the vertex boxes, to be moved by the nudging procedure in order to
maintain a minimum object distance \deltamin{} and to control the total
edge length.  We allow vertex boxes to grow, if necessary, but not to shrink.
Therefore, we use position variables for all sides of vertex boxes and edge
segments instead of fixed positions.

In addition to the constraints from the constrained mode (brown arrows in
\cref{fig:nudging-d}), we introduce, for each vertex box~$b$ of
original width $w_b$ (as specified in the input) a separation
constraint $b^{\mathrm{R}} - b^{\mathrm{L}} \ge w_b$ where $b^{\mathrm{L}}$
and $b^{\mathrm{R}}$ are the position variables of the left and right
sides of~$b$, respectively.  For all arcs that have not been transitively
removed (see brown and pink arrows in \cref{fig:nudging-c}), we add
separation constraints with a constant distance of~\deltamin.

In order to establish a hierarchy in optimization, we weight our
objective by adding constant factors.  Let $W$ be the set of distance
variables, let $S_{\mathrm{H}}$ be the set of horizontal segments,
and let $B$ the set of vertex boxes.  We use the term $b^{\mathrm{R}} -
b^{\mathrm{L}}$ for the width of box~$b \in B$ and $s^{\mathrm{R}} -
s^{\mathrm{L}}$ for the length of %
segment~$s \in S_{\mathrm{H}}$.
Now we have our LP minimize the sum of the widths of the vertex boxes
and the lengths of the segments minus the sum of the distance
variables, that is,
\[
  2(|W|+|S_{\mathrm{H}}|) \left(\omega + \sum_{b \in B} (b^{\mathrm{R}} - b^{\mathrm{L}}) \right)~
  + 2 \sum_{s \in S_{\mathrm{H}}} (s^{\mathrm{R}} - s^{\mathrm{L}})~
  - \sum_{\delta \in W} \delta .
\]

\cref{fig:nudging-f} shows the result for the example depicted in \cref{fig:nudging-a}.
Multiple phases of nudging can be repeatedly applied to optimize compactness and edge lengths.
To get rid of unnecessary bends, we simply set the separation distance of constraints
between segments of the same path to zero.

\section{Experiments}
\label{sec:experiments}

\pgfplotscreateplotcyclelist{three-marks}{
  {solid,mark options={scale=.3,fill=PKdarkblue},mark=diamond*,PKdarkblue},
  {solid,mark options={scale=.25,fill=PKdarkgreen},mark=triangle*,PKdarkgreen},
  {solid,mark options={scale=.25,fill=PKdarkorange},mark=*,PKdarkorange},
}
\pgfplotscreateplotcyclelist{semi-marks}{
  {draw opacity=0,fill opacity=0.2,mark options={scale=.4,fill=PKdarkblue},mark=diamond*,PKdarkblue},
  {solid,mark options={scale=.3,fill=PKdarkblue},mark=diamond*,PKdarkblue},
}

We considered three variants (described below) of our pipeline, and
compared them to the state-of-the-art orthogonal layout libraries
\textsc{Praline} and \textsc{Hola}.

\subsubsection{Benchmark Sets.}

Our pipeline has been implemented as part of a third-party-funded project with
two industrial partners that suggested two benchmark sets from their
respective domains.  The first benchmark set is called
\emph{Internet Topology Zoo}\footnote{see
\url{http://www.topology-zoo.org/index.html}}~\cite{Knight2011}.
The data set includes textual vertex labels of varying length.

The second dataset is called \emph{Pseudo-cableplans}.  %
The graphs have been part of a benchmark set for orthogonal graph drawing
by Zink et al.\footnote{see
\url{https://github.com/j-zink-wuerzburg/pseudo-praline-plan-generation}
}~\cite{zwbw-ldugg-CGTA22}.  We removed some
domain-specific peculiarities such as special vertex pairing and port
grouping constraints, and we replaced each hyperedge~$e$ by a new dummy
vertex~$v_e$ that we connected to every vertex in~$e$.  The labels in this
dataset are fixed-length or empty (the dummy vertices).

In both benchmark sets, we kept only the largest connected component
of each graph.  Although preliminary tests showed some good results,
\textsc{Hola}
officially does not support multigraphs.  Therefore, we removed all but
the first occurrence of each multi-edge and all self-loops.  Furthermore,
we removed all graphs where \textsc{Hola} crashed or took more than 10
minutes.  Note that our pipeline correctly draws every connected
instance in the original datasets, including multi-edges and self-loops.

\begin{figure}[tb]
  \centering
  \begin{subfigure}[b]{.49\linewidth}
    \centering
    \begin{tikzpicture}
      \begin{axis}[%
          width=.8\linewidth,
          height=.175\textheight,
          scale only axis,
          font=\small,
          cycle list name=semi-marks,
          only marks,
          minor tick num=1,
          ymax=450,
          xmax=260,
          xlabel={number of vertices},
          ylabel={number of edges},
        ]
        \addplot+ table[x=nodes,y=multi] {figures/tz-stats.dat};
        \addplot+ table[x=nodes,y=edges] {figures/tz-stats.dat};
      \end{axis}
    \end{tikzpicture}
    \subcaption{Internet Topology Zoo}
    \label{fig:tz-stats}
  \end{subfigure}
  \hfill
  \begin{subfigure}[b]{.49\linewidth}
    \centering
    \begin{tikzpicture}
      \begin{axis}[%
          width=.8\linewidth,
          height=.175\textheight,
          scale only axis,
          font=\small,
          cycle list name=semi-marks,
          only marks,
          minor tick num=1,
          ymax=450,
          xmax=260,
          xlabel={number of vertices},
        ]
        \addplot+ table[x=nodes,y=multi] {figures/pp-stats.dat};
        \addplot+ table[x=nodes,y=edges] {figures/pp-stats.dat};
      \end{axis}
    \end{tikzpicture}
    \subcaption{Pseudo-cableplans}
    \label{fig:pp-stats}
  \end{subfigure}
  \caption{Number of vertices and edges for each graph in the two datasets.
  Semi-transparent markers represent the original multigraphs.}
  \label{fig:dataset-stats}
\end{figure}
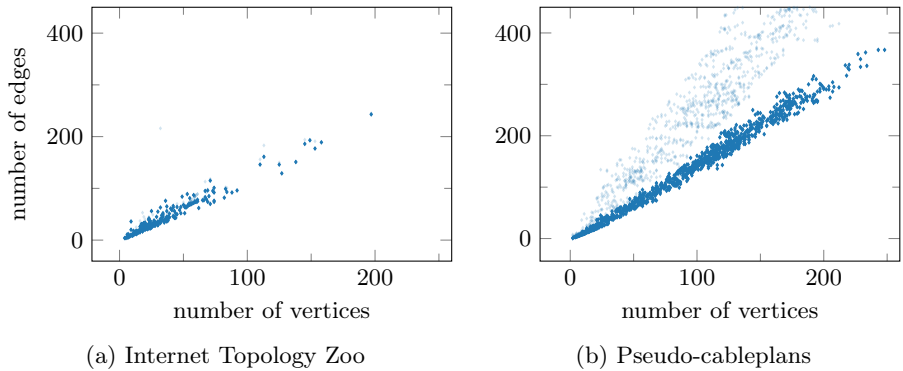

So for our experiments we used $260$ simple graphs derived from the original
$261$ multigraphs of the Internet Topology Zoo and 1,026 graphs derived from
the original 1,139 Pseudo-cableplans.  \Cref{fig:dataset-stats} shows the
edge density distribution of the graphs in the two benchmark sets.  We
set the default dimensions of the vertex boxes to $12\times38$
(pixels) and widened them if necessary to accommodate the label text
and to fit all incident edges (with gaps of 18 pixels).

\subsubsection{Pipeline Variants.}

We set up three variants of our pipeline.  In the first variant,
\emph{\textsc{Force}}, we use a simple force-directed layout
algorithm~\cite{fr-gdfdp-SPE91} to place the vertices as points
(ignoring their boxes).  Then, we apply the \mbox{\textsc{GTree}} Algorithm
by Nachmanson et al.~\cite{nachmanson2017} to remove overlaps.
In the second variant, \emph{\textsc{Hybrid1}}, we use vertex positions
computed by \textsc{Praline}~\cite{zwbw-ldugg-CGTA22}.
These variants both go through the steps port distribution,
routing graph construction, edge routing, edge ordering, and full edge nudging
as described in the previous sections.  Full nudging is applied horizontally,
then vertically, and then once more horizontally.
As a third variant \emph{\textsc{Hybrid2}}, we initialize our pipeline with
both the vertex positions and the edge routing computed by Praline and apply
only the nudging step as a post-processing.
All pipeline steps are implemented in Scala and dynamically configurable
for various setups.  We use the GLOP\footnote{see
\url{https://developers.google.com/optimization}}%
optimizer for LP-solving.

\subsubsection{Metrics.}

To assess the quality of graph drawings many metrics have been proposed.
In our experiments we use edge crossings, edge bends, total edge length,
variance in edge length, area, aspect ratio, and minimum object
distance (\deltamin). These will be discussed below.

It has been shown (e.g., in a study by Purchase~\cite{purchase2000})
that drawings with fewer edge crossings and bends simplify several
tasks related to graph understanding and navigation.  A study by Dwyer
et al.~\cite{dwyer2009} suggests that users benefit from graph
drawings with low variance in edge length.
When drawing graphs with more than 30 vertices, scaling becomes an issue
as text labels tend to become unreadably small and overly long edges become hard to follow.
Therefore, we include metrics assessing the compactness of drawings
in our comparisons, namely total edge length and the area of the bounding box.
For a drawing with a bounding box of width~$w$ and height~$h$,
we define aspect ratio as $\max(w, h)/\min(w, h)$.
This yields a value in the range $[1, \infty)$.  We consider lower
aspect ratios better and squares (with aspect ratio $1$) optimal.
In order to ensure a fair comparison with metrics sensitive to scaling,
we also include the minimum object distance \deltamin.
We configure a minimum value (or target value, if no minimum is supported) of
$12$\,pixels and report deviations.

\subsubsection{Comparison.}

We compared our pipeline to the following two libraries.
\textsc{Praline}~\cite{zwbw-ldugg-CGTA22} is
based on the well-known Sugiyama framework~\cite{Sugiyama1981}
for layered graph drawing.  \textsc{Praline} differs from the original
framework especially in terms of edge routing and port placement.
Layering-based algorithms tend to produce few crossings and
balanced results.
\textsc{Hola}~\cite{kieffer2016} is a multi-stage algorithm
that decomposes the input into trees and a connected core
that is drawn using stress-minimization and overlap removal.
The trees are drawn using a specialized layout algorithm, and the
tree drawings are then inserted into the drawing of the whole graph.
For our comparison, we used the default settings regarding vertex
distances and ideal edge length in \textsc{Hola}.  We conducted small-scale
experiments that confirm that the defaults yield a good compromise between
compactness and readability (i.e., sufficiently large \deltamin).

\subsubsection{Results.}

See \cref{tab:metrics} and \cref{fig:selected-metrics} for the results of our experiments.
Concerning the number of crossings, we see a weakness of the simplistic
approach of our pipeline.  On average, \textsc{Force} produced over twice as
many crossings as the best results on the Pseudo-cableplans and nearly five
times as many on the Internet Topology Zoo graphs.
\textsc{Hybrid1}, combining \textsc{Praline} vertex positions and our pipeline,
on the other hand, produced only $41\,\%$ more crossings on the
Pseudo-cableplans and $54\,\%$ more crossings on the Internet Topology Zoo graphs
compared to the best results.  \textsc{Hybrid2} per construction produces the
same number of crossings as \textsc{Praline}.
Bad vertex placement also hurts down the pipeline.
As we can see, the \textsc{Hybrid} variants are almost consistently better than
\textsc{Force} with two exceptions:  edge length variance and aspect ratio.
However, in terms of aspect ratio, the only outlier is \textsc{Hola},
performing nearly 30\,\% worse than the others on the Internet Topology Zoo.

\textsc{Hola} shows an impressive performance in terms of crossings and
creates by far the fewest bends.  But this comes at a cost of a very large
drawing area and overly long edges.  \textsc{Hola} considers \deltamin\ an
optimization goal, not a strict requirement.  In the majority of cases the
target value of 12\,px could be maintained.  \textsc{Praline}, however,
surprised us, too.  Not maintaining \deltamin\ was confirmed to us being a bug
in the current \textsc{Praline} implementation by the authors.

Overall, the \textsc{Hybrid} variants of our pipeline show good and
very consistent results with \textsc{Hybrid2}, surpassing \textsc{Praline}
in all quality metrics.  It produces leading results with respect to
area, total edge length, and edge length variance while reliably
maintaining the given minimum object distance.

\begin{table}[tb]
  \caption{Experimental results on two datasets.
  The mean $\mu$ is relative to \textsc{Praline} (abbreviated \textsc{Pral});
  $\beta$ measures the percentage of cases where an algorithm achieved the best result.
  Sums over 100\,\% are possible due to ties.}
  \label{tab:metrics}
  \medskip
  \setlength{\tabcolsep}{5pt}
  \newcolumntype{d}{D{.}{.}{1.2}}
  \centering
  \begin{subtable}{\textwidth}
  \caption{The Internet Topology Zoo benchmark set.}
  \centering
  \begin{tabular}{l drdrdrdrrr }
  \toprule
  & \multicolumn{2}{c}{\textsc{Force}} & \multicolumn{2}{c}{\textsc{Hybrid1}} & \multicolumn{2}{c}{\textsc{Hybrid2}} & \multicolumn{2}{c}{\textsc{Hola}} & \multicolumn{2}{c}{\textsc{Pral.}} \\
  & \multicolumn{1}{r}{$\mu$} & \multicolumn{1}{r}{$\beta$}
  & \multicolumn{1}{r}{$\mu$} & \multicolumn{1}{r}{$\beta$}
  & \multicolumn{1}{r}{$\mu$} & \multicolumn{1}{r}{$\beta$}
  & \multicolumn{1}{r}{$\mu$} & \multicolumn{1}{r}{$\beta$}
  & \multicolumn{1}{r}{~$\mu$} & \multicolumn{1}{r}{$\beta$} \\
  \midrule
    crossings     & 3.90 &      27  & 1.30 &      54  & 1.00 &      70  &  .85 & {\bf 77} & 1 &      70 \\
    edge\ bends   &  .92 &       4  &  .76 &       6  &  .49 & {\bf 51} &  .50 &      47  & 1 &       2 \\
    edge\ length\ variance
                  &  .47 & {\bf 39} &  .39 &      21  &  .37 &      38  &11.43 &       3  & 1 &       1 \\
    total\ edge\ length
                  &  .73 &      20  &  .61 &       6  &  .49 & {\bf 73} & 1.83 &       0  & 1 &       2 \\
    bounding\ box\ area
                  &  .68 &      30  &  .56 &      32  &  .56 & {\bf 38} & 3.94 &       0  & 1 &       0 \\
    aspect ratio  &  .93 & {\bf 35} & 1.03 &      14  & 1.07 &      11  & 1.35 &      22  & 1 &      18 \\
    \deltamin     & 1.11 & {\bf 88} & 1.10 &      87  & 1.10 &      87  & 1.01 &      45  & 1 &      69 \\
  \bottomrule
  \end{tabular}
  \end{subtable}
  \begin{subtable}{\textwidth}
  \caption{The Pseudo-cableplans benchmark set.}
  \centering
  \begin{tabular}{l drdrdrdrrr }
  \toprule
  & \multicolumn{2}{c}{\textsc{Force}} & \multicolumn{2}{c}{\textsc{Hybrid1}} & \multicolumn{2}{c}{\textsc{Hybrid2}} & \multicolumn{2}{c}{\textsc{Hola}} & \multicolumn{2}{c}{\textsc{Pral.}} \\
  & \multicolumn{1}{r}{$\mu$} & \multicolumn{1}{r}{$\beta$}
  & \multicolumn{1}{r}{$\mu$} & \multicolumn{1}{r}{$\beta$}
  & \multicolumn{1}{r}{$\mu$} & \multicolumn{1}{r}{$\beta$}
  & \multicolumn{1}{r}{$\mu$} & \multicolumn{1}{r}{$\beta$}
  & \multicolumn{1}{r}{~$\mu$} & \multicolumn{1}{r}{$\beta$} \\
  \midrule
    crossings     & 1.58 &       9  & 1.03 &      19  & 1.00 &      25  &  .73 & {\bf 89} & 1 &      25 \\
    edge\ bends   &  .96 &       1  &  .76 &       2  &  .66 &      12  &  .30 & {\bf 88} & 1 &       1 \\
    edge\ length\ variance
                  &  .34 & {\bf 52} &  .35 &      31  &  .40 &      19  & 1.79 &       0  & 1 &       0 \\
    total\ edge\ length
                  &  .59 &      29  &  .52 &      29  &  .54 & {\bf 43} & 1.20 &       0  & 1 &       0 \\
    bounding\ box\ area
                  &  .55 &      25  &  .50 &      24  &  .50 & {\bf 51} & 2.21 &       0  & 1 &       0 \\
    aspect ratio  &  .99 & {\bf 37} &  .97 &      12  &  .96 &      17  &  .97 &      24  & 1 &      10 \\
    \deltamin     & 1.42 & {\bf 93} & 1.42 & {\bf 93} & 1.42 & {\bf 93} & 1.24 &      60  & 1 &      33 \\
  \bottomrule
  \end{tabular}
  \end{subtable}
\end{table}

\begin{figure}[tb]
  \centering
  \begin{subfigure}[b]{.49\linewidth}
    \centering
    \begin{tikzpicture}
      \begin{semilogyaxis}[%
          width=.8\linewidth,
          height=.175\textheight,
          scale only axis,
          font=\small,
          cycle list name=three-marks,
          only marks,
          minor tick num=1,
          ymin=0.01,ymax=10,
          xmin=-10,xmax=230,
          xlabel={number of vertices},
          ylabel={number of edge bends},
        ]
        \draw[dotted] (-10,1) -- (230,1);
        \addplot+ table[x=n,y=hola] {figures/bends-tz.dat};
        \addplot+ table[x=n,y=y+]   {figures/bends-tz.dat};
      \end{semilogyaxis}
    \end{tikzpicture}
    \subcaption{Edge bends: Internet Topology Zoo}
    \label{fig:tz-bends}
  \end{subfigure}
  \hfill
  \begin{subfigure}[b]{.49\linewidth}
    \centering
    \begin{tikzpicture}
      \begin{semilogyaxis}[%
          width=.8\linewidth,
          height=.175\textheight,
          scale only axis,
          font=\small,
          only marks,
          cycle list name=three-marks,
          minor tick num=1,
          ymin=0.01,ymax=10,
          xmin=-10,xmax=230,
          xlabel={number of vertices},
          legend image post style={scale=4},
        ]
        \draw[dotted] (-10,1) -- (230,1);
        \addplot+ table[x=n,y=hola] {figures/bends-pp.dat};
        \addplot+ table[x=n,y=y+]   {figures/bends-pp.dat};
      \end{semilogyaxis}
    \end{tikzpicture}
    \subcaption{Edge bends: Pseudo-cableplans}
    \label{fig:pp-bends}
  \end{subfigure}

  \bigskip

  \begin{subfigure}[b]{.49\linewidth}
    \centering
    \begin{tikzpicture}
      \begin{semilogyaxis}[%
          width=.8\linewidth,
          height=.175\textheight,
          scale only axis,
          font=\small,
          cycle list name=three-marks,
          only marks,
          minor tick num=1,
          ymin=0.1,ymax=10,
          xmin=-10,xmax=230,
          xlabel={number of vertices},
          ylabel={bounding box area},
        ]
        \draw[dotted] (-10,1) -- (230,1);
        \addplot+ table[x=n,y=hola] {figures/area-tz.dat};
        \addplot+ table[x=n,y=y+]   {figures/area-tz.dat};
      \end{semilogyaxis}
    \end{tikzpicture}
    \subcaption{Area: Internet Topology Zoo}
    \label{fig:tz-area}
  \end{subfigure}
  \hfill
  \begin{subfigure}[b]{.49\linewidth}
    \centering
    \begin{tikzpicture}
      \begin{semilogyaxis}[%
          width=.8\linewidth,
          height=.175\textheight,
          scale only axis,
          font=\small,
          cycle list name=three-marks,
          minor tick num=1,
          ymin=0.1,ymax=10,
          xmin=-10,xmax=230,
          xlabel={number of vertices},
          legend image post style={scale=4},
        ]
        \draw[dotted] (-10,1) -- (230,1);
        \addplot+[only marks] table[x=n,y=hola] {figures/area-pp.dat}; \label{pgflabel:m-hola}
        \addplot+[only marks] table[x=n,y=y+]   {figures/area-pp.dat}; \label{pgflabel:m-hybr}
      \end{semilogyaxis}
    \end{tikzpicture}
    \subcaption{Area: Pseudo-cableplans}
    \label{fig:pp-area}
  \end{subfigure}
  \caption{
    Selected metrics of \textsc{Hola} \ref*{pgflabel:m-hola}, and \textsc{Hybrid2} \ref*{pgflabel:m-hybr}
    relative to \textsc{Praline}.}
  \label{fig:selected-metrics}
\end{figure}

\subsubsection{Running Time.}

We evaluated the running times of our pipeline using $300$ random
multigraphs with $5$ to $150$ vertices and average vertex degree~$4$.
To ensure that every graph is connected, we first created a tree with
all vertices and then added the remaining edges at random.
Our experiments ran on an %
Intel\textsuperscript{®} Core™ i7-8565U.  We measured runtimes using
Java's \texttt{nanoTime} function (for our pipeline and
\textsc{Praline}) and the GNU \texttt{time} command (for
\textsc{Hola}).

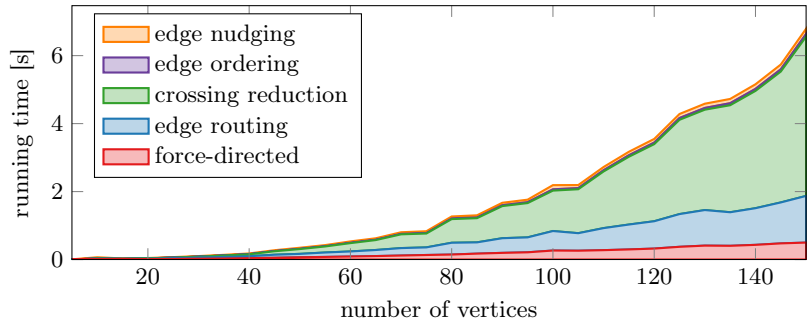
\begin{figure}[tb]
  \centering
  \begin{tikzpicture}
    \begin{axis}[%
        width=.8\linewidth,
        height=.175\textheight,
        scale only axis,
        font=\small,
        scaled y ticks=base 10:-3,
        ytick scale label code/.code={},
        xlabel={number of vertices},
        ylabel={running time [s]},
        stack plots=y,
        area style,
        enlarge x limits=false,
        enlarge y limits=upper,
        legend pos=north west,
        legend cell align=left,
        reverse legend=true,
      ]
      \addplot[thick,PKdarkred,fill=PKdarkred!30!white] table[x=n,y=fd] {figures/runtimes.dat} \closedcycle;
      \addplot[thick,PKdarkblue,fill=PKdarkblue!30!white] table[x=n,y=route] {figures/runtimes.dat} \closedcycle;
      \addplot[thick,PKdarkgreen,fill=PKdarkgreen!30!white] table[x=n,y=eyes] {figures/runtimes.dat} \closedcycle;
      \addplot[thick,PKdarkpurple,fill=PKdarkpurple!30!white] table[x=n,y=order] {figures/runtimes.dat} \closedcycle;
      \addplot[thick,PKdarkorange,fill=PKdarkorange!30!white] table[x=n,y=nudge] {figures/runtimes.dat} \closedcycle;

      \legend{force-directed,edge routing,crossing reduction,edge ordering,edge nudging}
    \end{axis}
  \end{tikzpicture}
  \caption{Running times of the \textsc{Force} pipeline on random multigraphs
    with an average vertex degree of $4$.  Stages with less than $10$\,ms
    average running time are omitted.}
  \label{fig:runtimes}
\end{figure}

See \cref{fig:runtimes} for different steps of our pipeline.
Steps that consistently require less than $10$\,ms to complete are omitted.
The overall runtime is clearly dominated by edge routing
and crossing reduction (that is, finding pairs of edges that cross
more than once and then joining their common subpaths),
followed by force-directed vertex layout.  The time spent on nudging,
edge ordering, and on creating the routing graph was insignificant.

The time for drawing the graphs from the two benchmark sets is shown in \cref{fig:dataset-runtimes}.
The \textsc{Hybrid} setups are omitted. In \textsc{Hybrid1} just like with \textsc{Force}
the edge routing dominates the runtime, in \textsc{Hybrid2} the runtime is dominated by performing
the \textsc{Praline} layout.  Note that \textsc{Praline} by default
does ten repetitions with different initial vertex positions of which it keeps the best.
Depicted is the sum of all repetitions.  For \textsc{Praline} and \textsc{Force}
only the bare layouting time was measured whereas for \textsc{Hola}, for technical
reasons, the measurements include file handling.
However, this increases the runtime by less than $50$\,ms.

\begin{figure}[tb]
  \centering
  \begin{subfigure}[b]{.49\linewidth}
    \centering
    \begin{tikzpicture}
      \begin{axis}[%
          width=.8\linewidth,
          height=.175\textheight,
          scale only axis,
          font=\small,
          cycle list name=three-marks,
          only marks,
          ymax=5.95,
          xmin=-10,xmax=230,
          xlabel={number of vertices},
          ylabel={running time [s]},
        ]
        \addplot+ table[x=nodes,y=hola]  {figures/runtimes-tz.dat};
        \addplot+ table[x=nodes,y=force] {figures/runtimes-tz.dat};
        \addplot+ table[x=nodes,y=pral]  {figures/runtimes-tz.dat};
      \end{axis}
    \end{tikzpicture}
    \subcaption{Internet Topology Zoo}
    \label{fig:tz-runtimes}
  \end{subfigure}
  \hfill
  \begin{subfigure}[b]{.49\linewidth}
    \centering
    \begin{tikzpicture}
      \begin{axis}[%
          width=.8\linewidth,
          height=.175\textheight,
          scale only axis,
          font=\small,
          cycle list name=three-marks,
          only marks,
          ymax=5.95,
          xmin=-10,xmax=230,
          xlabel={number of vertices},
          legend image post style={scale=4},
        ]
        \addplot+[only marks] table[x=nodes,y=hola]  {figures/runtimes-pp.dat}; \label{pgflabel:rt-hola}
        \addplot+[only marks] table[x=nodes,y=force] {figures/runtimes-pp.dat}; \label{pgflabel:rt-force}
        \addplot+[only marks] table[x=nodes,y=pral]  {figures/runtimes-pp.dat}; \label{pgflabel:rt-pral}
      \end{axis}
    \end{tikzpicture}
    \subcaption{Pseudo-cableplans}
    \label{fig:pp-runtimes}
  \end{subfigure}
  \caption{
    Running times of \textsc{Force} \ref*{pgflabel:rt-force},
    \textsc{Hola} \ref*{pgflabel:rt-hola}, and \textsc{Praline} \ref*{pgflabel:rt-pral}
    for our benchmark sets.}
  \label{fig:dataset-runtimes}
\end{figure}

\begin{figure}[p]
  \setlength{\fboxsep}{1pt}
  \centering
  \begin{minipage}{.45\linewidth}
  \begin{subfigure}[c]{\linewidth}
    \fbox{\includegraphics[scale=0.28]{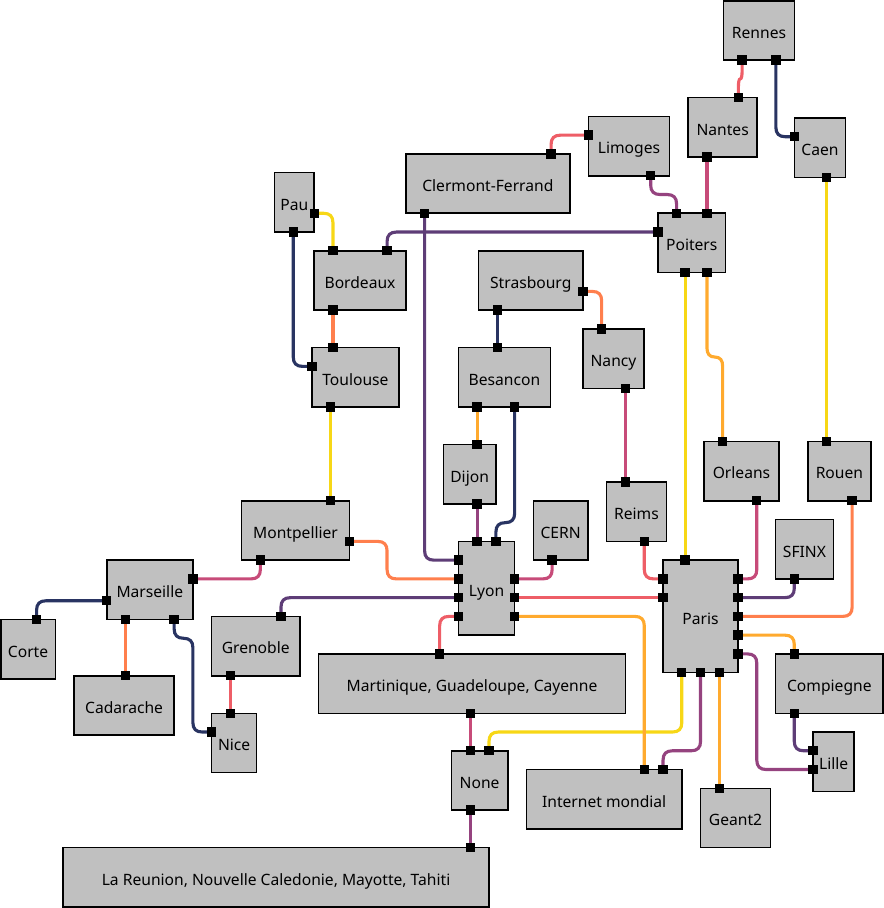}}
  \end{subfigure}
  \subcaption{\textsc{Hybrid1}: 2, 38, 12, 67\,\%
    \label{fig:example-hybrid1}}
  \vspace{5pt}
  \begin{subfigure}[c]{\linewidth}
    \fbox{\includegraphics[scale=0.28]{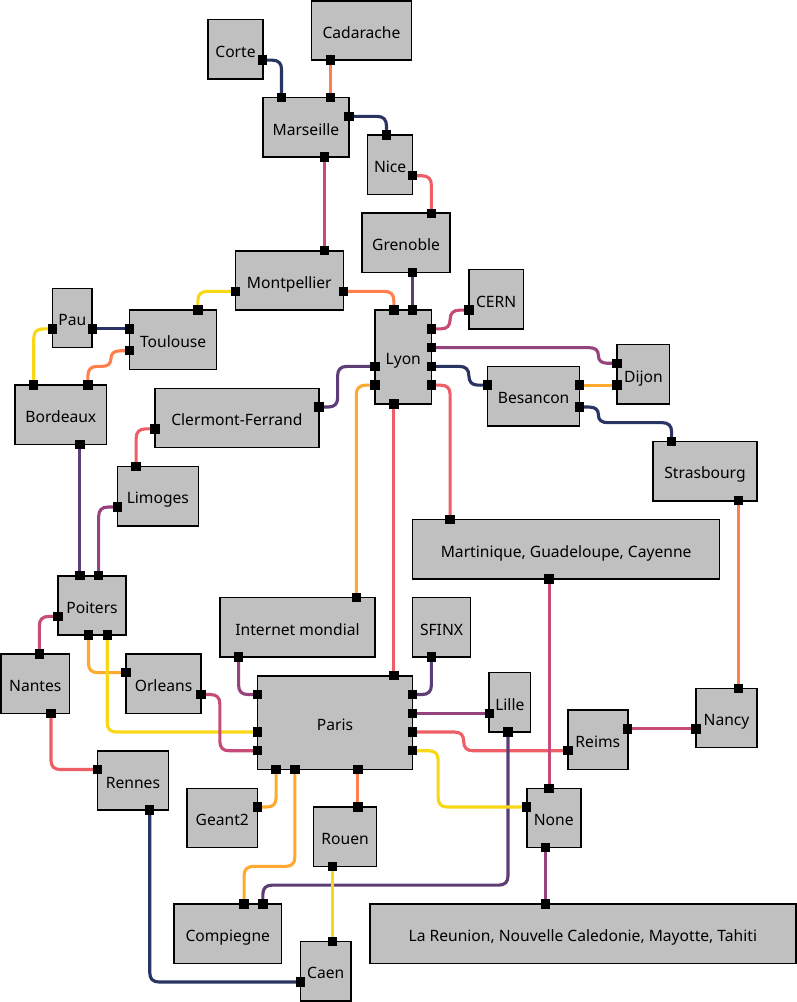}}
  \end{subfigure}
  \subcaption{\textsc{Force}: 6, 42, 12, 66\,\%
    \label{fig:example-force}}
  \vspace{5pt}
  \end{minipage}
  \begin{minipage}{.45\linewidth}
  \begin{subfigure}[c]{\linewidth}
    \fbox{\includegraphics[scale=0.28]{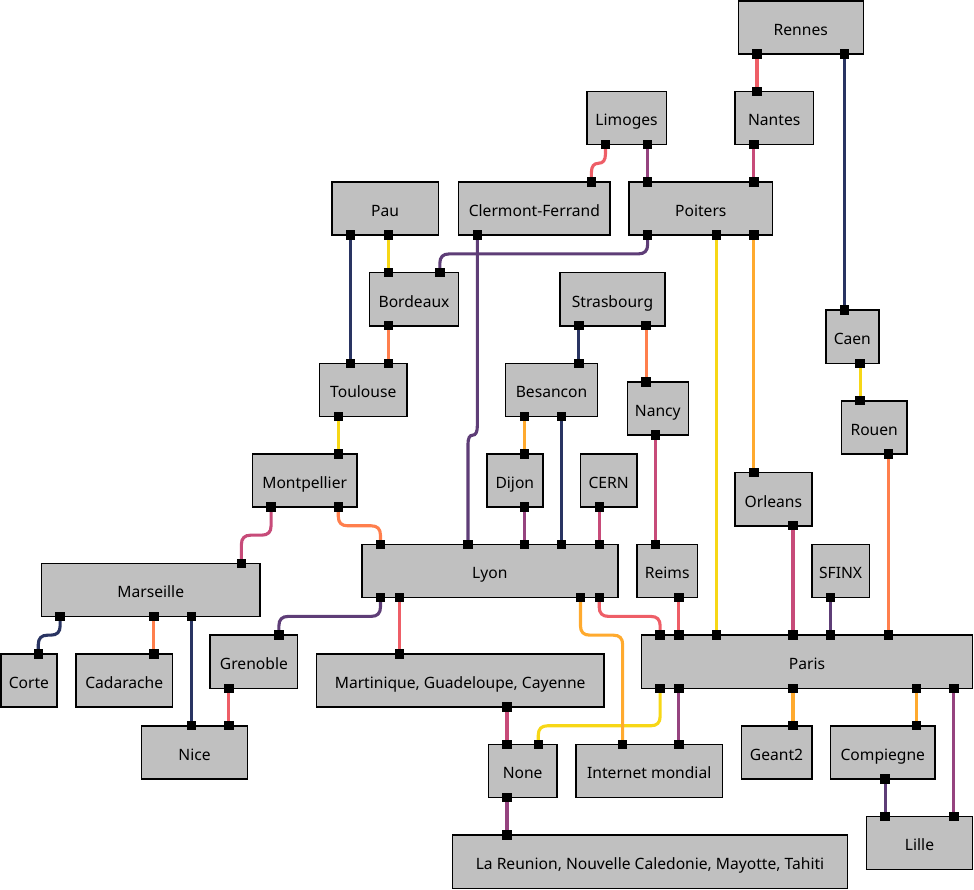}}
  \end{subfigure}
  \subcaption{\textsc{Hybrid2}: 2, 20, 12, 72\,\%
    \label{fig:example-hybrid2}}
  \vspace{5pt}
  \begin{subfigure}[c]{.54\linewidth}
    \fbox{\includegraphics[scale=0.28]{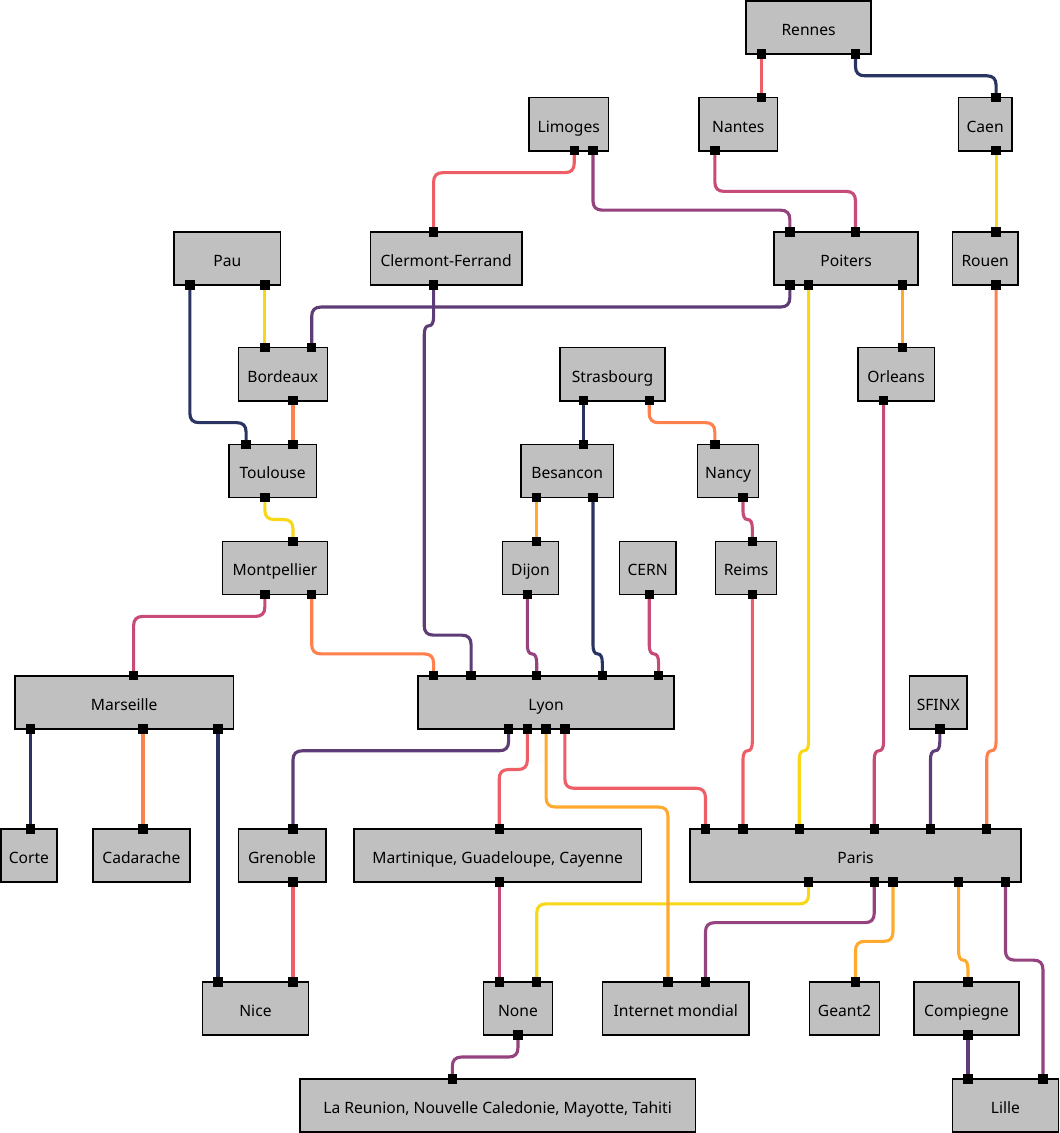}}
  \end{subfigure}
  \subcaption{\textsc{Praline}: 2, 62, 12, 100\,\%
    \label{fig:example-praline}}
  \vspace{5pt}
  \end{minipage}

  \begin{subfigure}[b]{\linewidth}
    \centering
    \fbox{\includegraphics[scale=0.28]{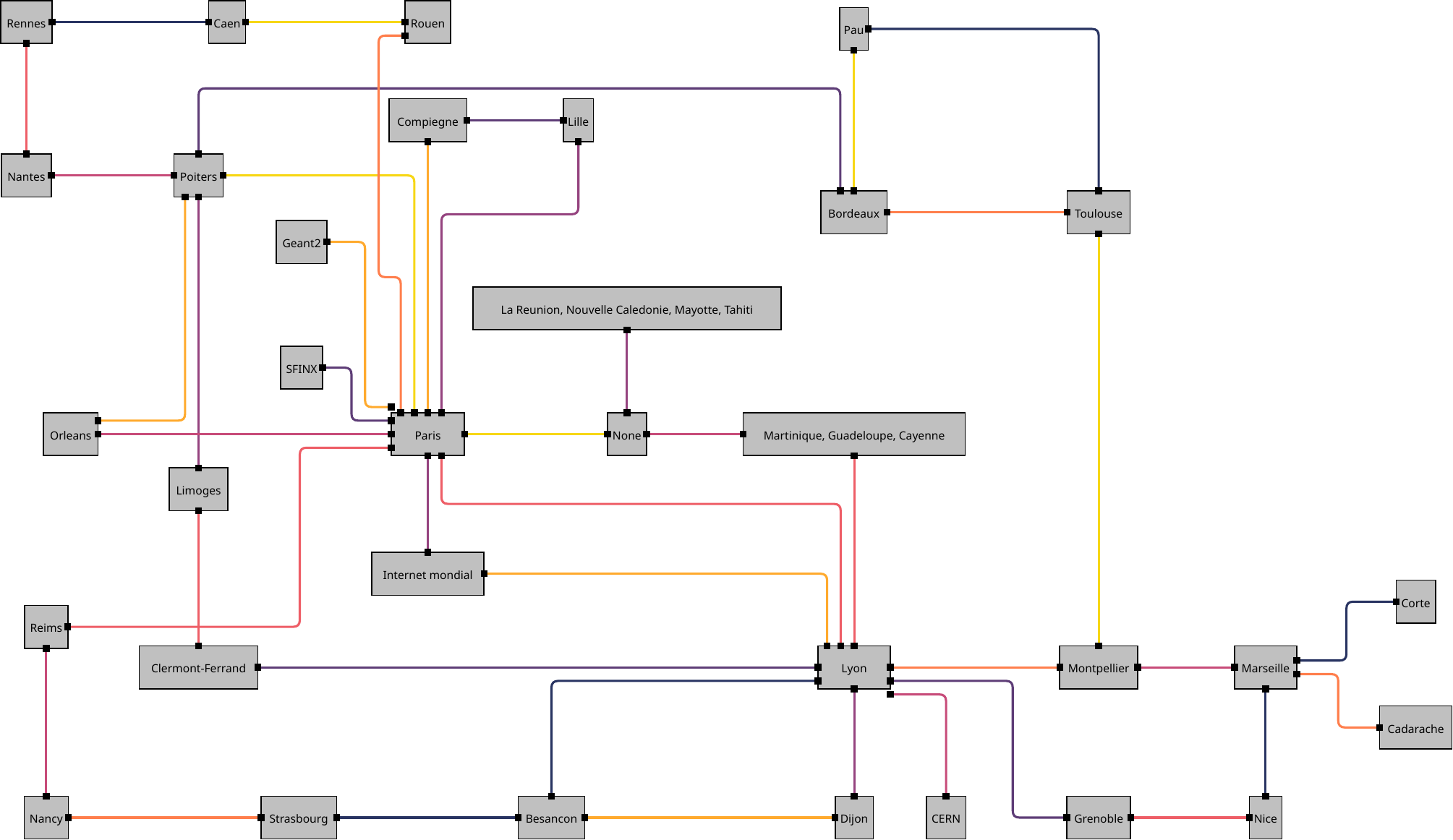}}
    \subcaption{\textsc{Hola}: 4, 27, 9, 195\,\%}
    \label{fig:example-hola}
  \end{subfigure}
  \caption{An example from the Internet Topology Zoo drawn by the five layout algorithms.
  The figures are scaled proportionally.  The numbers refer to: crossings,
  bends, $\deltamin$ (in pixels), and area (in percent w.r.t.\ \textsc{Praline}).}
  \label{fig:examples}
\end{figure}

\section{Conclusions}
\label{sec:conclusions}

Our experiments show that \textsc{Hybrid1} is a good allrounder.  The
edge nudging step performs particularly well and leads to compact
drawings.  On the other hand, due to our current rather simple edge routing,
the drawings tend to have more bends and crossings than those of its competitors.
When combined with a more sophisticated layouting (the \textsc{Hybrid2} setup),
we can significantly improve compactness (almost half the bounding box area)
and number of edge bends with the same number of crossings as \textsc{Praline}.

Nonetheless, we intend to improve edge routing, especially in terms of %
crossings.  To this end, Wybrow et al.~\cite{wybrow2010}
suggested to take into account edges that have already been routed.
Also it may help to reorder the ports around the boundary of the
vertex boxes.  %
To reduce the number of bends, we want to add a postprocessing that
straightens Z-shaped edges whose middle piece is short.
Currently, such unnecessary double bends tend to occur quite
frequently; see \cref{fig:example-force}.

\subsubsection{Acknowledgments.}

We thank Steve Kieffer, Micheal Wybrow, and Tobias Czauderna who
helped us with \textsc{Hola} in our experiments, Johannes Zink who
helped us with \textsc{Praline}, and our very supportive reviewers.
This work was supported by BMBF grant 01IS22012C.

\bibliographystyle{abbrvurl}
\bibliography{abbrv,wueortho,orthogonal-graph-drawing}

\end{document}